%% file: skeleton-cover-test.tex
\documentclass[11pt]{article}

\usepackage{amsfonts,epsfig,fullpage}
\usepackage{amsmath}
\usepackage[table]{xcolor}
\usepackage{caption}
\usepackage{graphicx}

\include{pream}
\newcommand{\Z}{{Z}}

\definecolor{awesome}{rgb}{1.0, 0.13, 0.32}
\definecolor{crimsonglory}{rgb}{0.75, 0.0, 0.2}
\definecolor{blue-violet}{rgb}{0.54, 0.17, 0.89}
\definecolor{darkmagenta}{rgb}{0.55, 0.0, 0.55}
\definecolor{electriccrimson}{rgb}{1.0, 0.0, 0.25}
\definecolor{scarlet}{rgb}{1.0, 0.13, 0.0}
\definecolor{indigo(web)}{rgb}{0.29, 0.0, 0.51}
\definecolor{burgundy}{rgb}{0.5, 0.0, 0.13}

\newcommand{\dcom}[1]{{\color{blue}{D: #1}}}
\newcommand{\mcom}[1]{{\color{purple}{M: #1}}}
\newcommand{\mchange}[1]{{\color{darkmagenta}  #1}}

\begin{document}

\title{Property testing of the  Boolean and binary rank}
\author{
    Michal Parnas \\
    The Academic College\\
    of Tel-Aviv-Yaffo \\
     {\tt michalp@mta.ac.il}
\and
 Dana Ron \\
 Tel-Aviv University,\\
 {\tt danaron@tau.ac.il}
\and
    Adi Shraibman \\
   The Academic College\\
    of Tel-Aviv-Yaffo \\
     {\tt adish@mta.ac.il}
}

\maketitle

\abstract{
We present algorithms for testing if a $(0,1)$-matrix $M$ has Boolean/binary rank at most $d$,
or is $\epsilon$-far from Boolean/binary rank $d$
(i.e., at least an $\epsilon$-fraction of the entries in $M$ must be modified so that it has rank at most $d$).

The query complexity of our testing algorithm for the Boolean rank is $\tilde{O}\left(d^4/ \epsilon^6\right)$.
For the binary rank we present a testing algorithm whose query complexity is $O(2^{2d}/\epsilon)$.
Both algorithms are $1$-sided error algorithms that always accept $M$ if it has Boolean/binary rank at most $d$,
and reject with probability at least $2/3$ if $M$ is $\epsilon$-far from Boolean/binary rank $d$.
}\\

\section{Introduction}

The Boolean rank of a $(0,1)$-matrix $M$ of size $n\times m$ is equal to the minimal $r$,
such that $M$ can be factorized as a product $M = X\cdot Y$,   where
$X$ is $(0,1)$-matrix of size  $n \times r$
and $Y$ is a  $(0,1)$-matrix of size $r \times m$, and all additions and multiplications are Boolean (that is, $1+1 = 1, 1+0 = 0+1 = 1, 1\cdot 1 = 1$).
A similar definition holds for the binary rank, where here the operations are the regular operations over the reals (that is, $1+ 1 = 2$).

These two rank functions have other equivalent definitions:
The Boolean (binary) rank is equal to the minimal number of monochromatic rectangles required to cover (partition) all the $1$-entries of the matrix.
The Boolean (binary) rank is also equal to the minimal number of bipartite cliques needed to cover (partition) all the
edges of a bipartite graph whose adjacency matrix is $M$ (see~\cite{Gregory}).
Furthermore, the Boolean rank of $M$ determines exactly the non-deterministic communication complexity of $M$, and
the binary rank of $M$ gives an approximation up to a polynomial of the deterministic communication complexity of $M$
(see, for example,~\cite{KN97} for more details).

Given the importance of these two rank functions it is desirable to be able to compute or approximate them efficiently.
However,
in several works it was shown that computing and even approximating the Boolean or binary rank  is NP-hard~\cite{orlin1977contentment,simon1990approximate,gruber2007inapproximability,CHHK14}.
The strongest inapproximability result~\cite{CHHK14} shows that it is NP-hard to approximate both ranks  to within a factor of $n^{1-\delta}$ for any given $\delta >0$,
and using a stronger complexity assumption they prove a lower bound that is even closer to linear in $n$.

\subsection{Property testing of the matrix rank}  
In this work we consider a different relaxation of exactly computing the Boolean or binary rank of a matrix, namely, that of \emph{property testing}~\cite{RS96,GGR98}.
For a parameter $\eps \in [0,1]$ and an integer $d$, a matrix $M$ is said to be \emph{$\eps$-far} from Boolean (binary) rank at most $d$,
if it is necessary to modify more than an $\eps$-fraction of the entries of $M$ to obtain a matrix with Boolean (binary) rank at most $d$.
Otherwise, $M$ is \emph{$\eps$-close} to Boolean (binary) rank at most $d$.

A property-testing algorithm for the Boolean (binary) rank is given as parameters $\eps$ and $d$,
as well as query access to a matrix $M$. If $M$ has Boolean (binary) rank at most $d$, then the algorithm should accept with probability at least $2/3$,
and if $M$ is $\eps$-far from having Boolean (binary) rank at most $d$, then the algorithm should reject with probability at least $2/3$.
If the algorithm accepts matrices having rank at most $d$ with probability 1, then it is a \emph{one-sided error} testing algorithm.
If it selects all its queries in advance, the it is a \emph{non-adaptive} algorithm.
The main complexity measure that we focus on is the query complexity of the testing algorithm.



\paragraph{The real rank.}
If one considers the real rank of matrices, then there are known efficient property testing algorithms.
Krauthgamer and Sasson~\cite{krauthgamer2003property} gave a non-adaptive property testing algorithm for the real rank
whose query complexity is $O(d^2 / \epsilon^2)$, and Li, Wang and Woodruff~\cite{Li}
showed that by allowing the algorithm to be adaptive, it is possible to reduce the
query complexity to $O(d^2 / \epsilon)$.
Recently, Balcan et al.~\cite{BLWZ} gave a non-adaptive testing algorithm for
the real rank whose query complexity is $\tilde{O}(d^2 / \epsilon)$.

It should be noted that the aforementioned property testing algorithms  for the real rank cannot be simply adapted to the Boolean and binary rank,
since they rely heavily on the augmentation property that holds trivially for the real rank
(i.e., if the real rank of $(M|x)$ and of  $(M|y)$ is $d$, then the real rank of $(M|x,y)$ is also $d$,
where $(M|x)$ is the matrix $M$ augmented with a vector $x$ as the last column).
However, the augmentation property does not hold for the Boolean and binary rank (see for example~\cite{parnas2018augmentation}),
and thus a different approach is needed.

\paragraph{Induced-subgraph freeness in bipartite graphs.}
Recalling the formulation of the Boolean and binary rank as properties of bipartite graphs,
we observe that these properties can be characterized as being free of a finite collection of induced subgraphs.
Alon, Fischer and Newman \cite{alon2007efficient}  showed that
every such property of bipartite graphs
can be tested with a number of queries that is polynomial in $1/ \epsilon$, and with no dependence on the size of the graph.

However, by applying their framework to our problems, we obtain algorithms whose complexity is
quite high as a function of $d$. It is not hard to verify, as we show for completeness in Section~\ref{sec:Alon},
that in the worst case, the query complexity achieved is upper bounded by $(\frac{2^d}{\epsilon})^{O(2^{4d})}$.

Specifically, we define the set $F_{d+1}$ of all $(0,1)$-matrices of Boolean (binary) rank $d+1$, without
repetitions of rows or columns in the matrices in $F_{d+1}$,
and then use the result of \cite{alon2007efficient} to test for matrices that do not contain as a submatrix any member of $F_{d+1}$.
The query complexity of the resulting algorithm depends on the maximal size of the matrices in $F_{d+1}$,
where this size is upper bounded by $2^{d+1}$.


\paragraph{Relation to graph coloring.}
The formulation of the Boolean (binary) rank as a covering (partition) problem of edges by complete bicliques
is reminiscent of the well-known problem of graph coloring, where the goal is to partition the graph vertices into a small number of independent sets.
Graph coloring has been studied in the context of property testing, where the query complexity is polynomial in $1/\eps$ and the number, $k$, of colors  (see~\cite{GGR98,AK02,Sohler12}).
However, an important difference between the problems is that while $k$-colorability is monotone in terms of the removal of edges,
this is not true of the Boolean and binary rank.
In particular, this implies that when testing $k$-colorability, the distance of a graph $G$ to the property is the minimal number of $1$-entries
in the adjacency matrix of $G$ that should be modified to $0$, so as to obtain a $k$-colorable graph.
On the other hand, for the rank functions we consider, if we want to make the minimal number of modifications in a $(0,1)$-matrix $M$,
so as to obtain a $(0,1)$-matrix  with rank at most $d$, then we might need to modify both $1$-entries and $0$-entries.

\subsection{Our results}

Our first and main result is a property testing algorithm for the Boolean rank that has query complexity polynomial  in $d$ and $1/\epsilon$.


\BT
\label{thm:main}
There exists a one-sided error non-adaptive property testing algorithm for the Boolean rank 
whose query complexity is $\tilde{O}\left(d^4/ \epsilon^6\right)$.
\ET

The proof of Theorem~\ref{thm:main} builds on the framework used in~\cite{parnas2006tolerant}, which in turn builds on~\cite{czumaj2005abstract}.
Specifically, we introduce the notion of {\em skeletons\/} and {\em beneficial entries\/} for a matrix $M$
in the context of the Boolean rank. These notions allow us to separate the analysis of Algorithm~\ref{alg:Boolean test} into a purely combinatorial part and a probabilistic part.
Part of the challenge in defining these notions in the context of the Boolean rank, and using them to prove Theorem~\ref{thm:main}, is the non-monotonicity of the problem described above.
More details on the proof structure, as well as the complete proof of Theorem~\ref{thm:main}, are given in Section~\ref{sec:Boolean}.

\smallskip
For the binary rank we present testing algorithms whose query complexity  is exponential in the rank $d$.

\BT
\label{thm:binary}
There exists a one-sided error non-adaptive property testing algorithm for the binary rank 
whose query complexity is $O(2^{2d}/\epsilon^2)$,
and a $1$-sided error adaptive property testing algorithm for the binary rank 
whose query complexity is $O(2^{2d}/\epsilon)$.
\ET

The proof of Theorem~\ref{thm:binary} is given in Section~\ref{sec:binary}.
Observe that even if it turns out that the property of having binary rank at most $d$ can be characterized
by being free of all submatrices that belong to a family $F$ of matrices having size $O(d)$,
still our algorithms are an improvement over the result that can be derived from~\cite{alon2007efficient},
which in this case would be $(d/\epsilon)^{O(d^4)} = 2^{O(d^4\log d)}/\epsilon^{O(d^4)}$ (see Section~\ref{sec:Alon} for more details).
It remains an open problem whether there exists a testing algorithm for the binary rank with  query complexity polynomial in $d$ and $1/\epsilon$.


\section{Testing the Boolean rank} 
\label{sec:Boolean}

Let $M$ be a $(0,1)$-matrix of size  $n\times n$, and let $[n] = \{1,\dots,n\}$.
We say that an entry $(x,y)\in [n]\times[n]$ is a \emph{$1$-entry} of $M$ if $M[x,y]=1$.
For a subset of entries $U = \{(x_i,y_i)\}_{i=1}^{m}$, the submatrix of $M$ \emph{induced} by
$U$ is the submatrix whose rows are $\{x_i\}_{i=1}^{m}$ and whose columns are
$\{y_i\}_{i=1}^{m}$. 
The testing algorithm for the Boolean rank is simple:

\bigskip
\fbox{
\begin{minipage}{5.4in}
\BA{{\sf(Test $M$ for Boolean rank $d$, given $d$ and $\eps$)}}
\label{alg:Boolean test}
\begin{enumerate}
\item
Select uniformly, 
independently and at random
$m = \Theta\left(\frac{d^2}{\eps^3}\cdot \log \frac{d}{\eps}\right)$
entries from $M$.
\item
Let $U$ be the subset of entries selected 
and consider the submatrix $W$ of $M$ induced by $U$.
\item
If $W$ has Boolean rank at most $d$, then accept. Otherwise, reject.
\end{enumerate}
\EA
\end{minipage}
}
\bigskip


The query complexity of Algorithm~\ref{alg:Boolean test}  is clearly $\tilde{O}\left(d^4/ \epsilon^6\right)$.
As 
for the running time of the algorithm, we cannot expect it to be efficient since computing the Boolean rank of a $(0,1)$-matrix is NP-hard.
We now proceed to prove Theorem~\ref{thm:main}, and show that Algorithm~\ref{alg:Boolean test} is a $1$-sided error tester for the Boolean rank.

\paragraph{Proof Structure.}
First note that if $M$ has Boolean rank at most $d$, then so does each of its submatrices,
causing Algorithm~\ref{alg:Boolean test} to accept (with probability 1). Hence,
our focus is on proving that if $M$ is $\eps$-far from Boolean rank $d$, then the algorithm rejects
with probability at least $2/3$.
Following~\cite{parnas2006tolerant}, which in turn follow~\cite{czumaj2005abstract},
we introduce the notion of {\em skeletons\/} and {\em beneficial entries\/} for a matrix $M$
in the context of the Boolean rank. These notions allow us to separate the analysis of Algorithm~\ref{alg:Boolean test} into
a purely combinatorial part (which is the main part of the analysis), and a probabilistic part (which is fairly simple).

A skeleton for $M$ is a multiset $S = \{S_1,\dots,S_d\}$ that contains  $d$ subsets of $1$-entries of $M$,
where all $1$-entries in each subset $S_i$ can be in the same monochromatic rectangle.
Roughly speaking, an entry $(x,y)$ is beneficial with respect to a skeleton  $S = \{S_1,\dots,S_d\}$, if for each one of the subsets $S_i$,
either: (1) $(x,y)$ cannot be added to $S_i$, since it cannot be in the same monochromatic rectangle as the entries already in $S_i$,
or (2) adding $(x,y)$ to $S_i$  significantly reduces the number of other entries that can be in the same monochromatic rectangle
with the entries of $S_i$ and $(x,y)$.

Observe that if an entry $(x,y)$ cannot be added to \emph{any} $S_i$ in $S$, then this is evidence that the skeleton $S$
cannot be extended to a cover of all $1$-entries of $M$ by monochromatic rectangles. More generally, allowing also for the second option
defined above, beneficial entries make a skeleton more \emph{constrained}, as we formalize precisely in the next subsection.

We show how, given a matrix $M$, it is possible to define a set $\calS(M)$ of relatively small skeletons that have certain useful properties.
In particular, if $M$ is $\epsilon$-far from Boolean rank at most $d$, then every skeleton in $\calS(M)$ has many beneficial entries.
We establish this claim by showing how, given a skeleton $S$ in $\calS(M)$, we can modify $M$ so as to obtain a matrix with Boolean rank at most $d$,
where the number of modifications is  upper bounded as a function of the number of entries that are beneficial with respect to $S$.
On the other hand, we show that if the matrix $M$ has Boolean rank at most $d$, then for every submatrix $W$ of $M$,
every subset of entries $U \subseteq W$ contains a skeleton in $\calS(M)$ with no beneficial entries in $U$.
Finally, we prove that if every skeleton in $\calS(M)$  has many beneficial entries, then with high constant probability
over the choice of $U$ in Algorithm~\ref{alg:Boolean test}, for every skeleton $S \in \calS(M)$ that is contained in $U$,
there exists a beneficial entry in $U$ for $S$. We note that the bound on the size of the skeletons in $\calS(M)$ plays a role in this last proof.
 Theorem~\ref{thm:main} can then be shown to follow by combining the above.

\subsection{Central definitions}
\label{subsec:cent-def}
Throughout this subsection, the matrix $M$ and the parameters $d$ and $\eps$ are fixed.

\BD[Compatible entries]
\label{def:compatible}
An entry $(x_1,y_1)$ is {\sf compatible} with a $1$-entry $(x_2,y_2)$ if $M[x_1,y_2] = M[x_2,y_1]=1$.
Otherwise, $(x_1,y_1)$ is {\sf incompatible} with $(x_2,y_2)$.

An entry $(x,y)$ is {\sf compatible with a set} $S$ of $1$-entries, if $(x,y)$ is compatible with every entry in $S$. Otherwise, $(x,y)$ is {\sf incompatible} with $S$.
\ED
Note that if both entries $(x_1,y_1)$ and $(x_2,y_2)$ are 1-entries, then 
the compatibility relation is symmetric,
but it applies also to pairs of entries $(x_1,y_1)$ and $(x_2,y_2)$ such that $M[x_1,y_1]=0$ and $M[x_2,y_2]=1$.
When both entries $(x_1,y_1)$ and $(x_2,y_2)$ are 1-entries, compatibility means that these entries can belong
to the same monochromatic rectangle.

\BD[Friendly row/column]
\label{def:friendly}
A row $x$ (column $y$) is {\sf friendly\/} with a set $S$ of $1$-entries,
if  for every entry $(x',y') \in S$, it holds that $M[x,y'] = 1$ ($M[x',y] = 1$).
Otherwise, it is {\sf not friendly} with $S$.
\ED

\begin{figure}[htb!]
\captionsetup{width=0.9\textwidth}
\centering
    \includegraphics[width=0.25\textwidth]{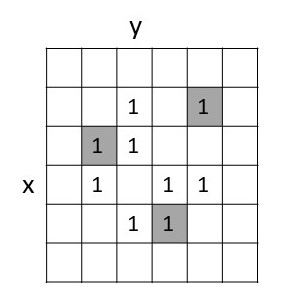}
\caption{\small Row $x$ and column $y$ are friendly with the subset $S$ of grey entries, and entry $(x,y)$ is compatible with $S$.
The empty entries can be either $1$ or $0$.}
\label{fig:friendly}
\end{figure}

Observe that by Definitions~\ref{def:compatible} and~\ref{def:friendly},
an entry $(x,y)$ is compatible with a set $S$ of $1$-entries  if and only if row $x$ and column $y$ are both friendly with $S$.
See Figure~\ref{fig:friendly} for an illustration.

For a set of entries $S = \{(x_i,y_i)\}_{i=1}^{|S|}$, denote the set of rows of $S$ by $R(S) = \{x_i\}_{i=1}^{|S|}$ and
the set of columns of $S$ by $C(S) = \{y_i\}_{i=1}^{|S|}$.

\BD[Zeros of a row/column]
For a row $x$, let $\Z(x)$ denote the set of columns $y$ such that
$M[x,y]=0$, and for a column $y$ let $\Z(y)$ denote the set of rows $x$ such that $M[x,y]=0$.

We extend the notation $\Z(\cdot)$  to sets of rows/columns.
Namely, for a set of rows $X$,  $\Z(X) = \displaystyle{\bigcup_{x\in X}}\Z(x)$,
and similarly for a set of columns $Y$.

\ED


\BCM
\label{clm:fewzeros}
Let $(x,y)$ be an entry such that $M[x,y] = 0$ and such that row $x$ and column $y$ are both friendly with
a set of entries $S$.
Then
$y \in \Z(x) \setminus \Z(R(S))$ and $x \in \Z(y) \setminus \Z(C(S))$.
\ECM
\BPF
Assume, contrary to the claim, that $y \notin \Z(x) \setminus \Z(R(S))$.
Since $M[x,y]=0$ then $y \in \Z(x)$, and therefore, $y \in \Z(R(S))$.
Hence, there exists an entry $(x',y')\in S$ such that $M[x',y] = 0$.
But this means that column $y$ is not friendly with $S$.

An analogous argument shows that $x \in \Z(y) \setminus \Z(C(S))$.
\EPF\\

\BD[Zero-heavy row/column]\label{def:zero-heavy}
Row $x$ is {\sf zero-heavy} with respect to a set of entries $S$ if
$|\Z(x) \setminus \Z(R(S))|
      \geq g(\eps,d)\cdot n$,
where $g(\eps,k) = \eps/(4d)$.
Otherwise, it is {\sf zero-light} with respect to $S$.
 Similarly,  column $y$ is zero-heavy with respect to $S$ if
$|\Z(y) \setminus \Z(C(S))|  \geq g(\eps,d)\cdot n$, and otherwise, it is zero-light.
\ED

\BD[Influential entries]
\label{def:influential}
Entry $(x,y)$ is {\sf influential} with respect to a set $S$ of $1$-entries if:
(1) $M[x,y] = 1$, (2)  $(x,y)$ is compatible with $S$,
and (3) either row $x$ or column $y$ is zero-heavy with respect to $S$ (possibly both).
Otherwise, $(x,y)$ is {\sf non-influential} for $S$.
\ED

\begin{figure}[htb!]
\captionsetup{width=0.9\textwidth}
\centering
    \includegraphics[width=0.5\textwidth]{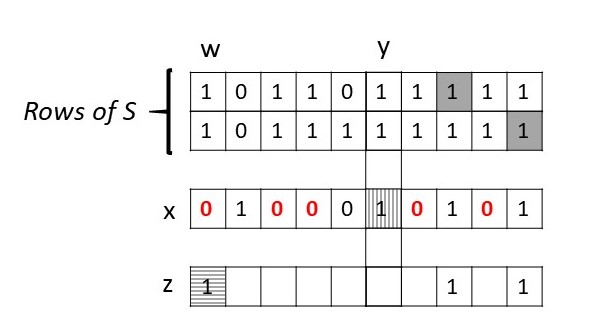}
  \caption{\small An illustration for the definition of a zero-heavy row and influential entries.
  The entries of $S$ are filled with grey.
  The 0-entries  in row $x$ that belong to $\Z(x) \setminus \Z(R(S))$ are colored red.
  Consider the $1$-entry $(x,y)$ that is filled with vertical lines. Assuming that row $x$ is zero-heavy, then entry $(x,y)$ is influential with respect to $S$.
  Furthermore, although the $1$-entry $(z,w)$, which is filled with horizontal lines, is compatible with $S$, it is not compatible with $S\cup \{(x,y)\}$.}
  \label{fig:zero-heavy}
\end{figure}

As we will see shortly, only influential entries will be added to a given skeleton.
This will allow us to maintain small skeletons and at the same time, when $M$ is $\epsilon$-far from Boolean rank $d$,
each skeleton will have many beneficial entries, as defined next.
An illustration for Definitions~\ref{def:zero-heavy} and~\ref{def:influential} is given in  Figure~\ref{fig:zero-heavy}.

We are now ready to introduce our main definitions of skeletons and beneficial entries.

\BD [Skeletons and beneficial entries for the Boolean rank]
\label{def:skeletons-and-benefical}
\sloppy
A {\sf skeleton} for a matrix $M$ is a multiset $S = \{S_1,\ldots,S_d\}$ that includes $d$ subsets of $1$-entries of $M$,
and  is defined inductively as follows:
\begin{enumerate}
\item
The multiset $S = \{\emptyset,\dots,\emptyset\}$,  
which contains the empty set $d$ times, is a skeleton.
\item
If $S=\{S_1,\ldots,S_d\}$ is a skeleton and $(x,y)$
is an influential entry  with respect to $S_i$ for some
$i \in [d]$,
then 
$S' =
   \{S_1,\ldots,S_{i-1},S_i \cup \{(x,y)\},S_{i+1},\ldots,S_d)\}$
is a skeleton.

(Note that there may be more than one way to add $(x,y)$ to the skeleton $S$, and $(x,y)$ can be added to more than one of the subsets $S_i$).
\end{enumerate}
Let $\calS(M)$ denote the set of all skeletons for $M$.

A $1$-entry $(x,y) \in M$ is {\sf beneficial} for a skeleton $S = \{S_1,\ldots,S_d\}$, if
for every $1 \leq i \leq d$, the entry $(x,y)$ is either incompatible or influential with respect to $S_i$.
Otherwise, $(x,y)$ is {\sf non-beneficial} for $S$.
\ED
Note that by the definition of influential entries, any skeleton $S\in \calS(M)$ contains only $1$-entries of $M$, and beneficial entries are always $1$-entries.

In the next two subsections we prove that the set of skeletons $\calS(M)$, as defined in Definition~\ref{def:skeletons-and-benefical},
has certain properties, which are then exploited to prove Theorem~\ref{thm:main}.

\subsection{Matrices of 
              rank  $d$ have  skeletons with no beneficial entries}
\label{subsec:rank-k-no-beneficial}

\BL\label{lem:rank-k-no-beneficial}
Let $W$ be a submatrix of $M$ with Boolean rank at most $d$.
Then for every $U \subseteq W$, there exists a skeleton $S =\{S_1,\ldots,S_d\}\in \calS(M)$, such that
$\bigcup_{i=1}^d S_i \subset U$, and there is no beneficial entry in $U$ for $S$.
\EL
\BPF
First observe that since $W$ has Boolean rank at most $d$, there exist
$d$ monochromatic submatrices $B_1,\dots,B_d$ of $W$ that cover all $1$-entries of $W$, and hence all $1$-entries of $U$.
We  build the skeleton $S$ in the following iterative manner:
\begin{enumerate}
\item
We start with the skeleton
$S^1 = \{\emptyset,\ldots,\emptyset\}$.
\item
Let $S^j = \{S^j_i\}_{i=1}^d$
be the skeleton at the beginning of the $j$'th iteration.
\begin{enumerate}
\item If there exists an index $i$ and an entry
$(x,y) \in B_i  \cap U$ that is an influential entry with respect to
$S^j_i$,
then we let
$S^{j+1} =
  \{S^j_1,\ldots,S^j_{i-1},S^j_i \cup \{(x,y)\},S^j_{i+1},\ldots,S^j_d\}$.
\item
If for every $i$, the subset 
$B_i \cap U$ does not
contain any influential entry with respect $S^j_i$, then we stop.
 \end{enumerate}
\end{enumerate}


Let $S=\{S_1,\dots,S_d\}$ be the final resulting skeleton.
It remains to show that there are no beneficial entries in 
$U$ for $S$.

Assume, contrary to this claim, that there is some beneficial $1$-entry 
$(x,y)\in U$ for $S$.
Since the submatrices $B_1,\dots,B_d$ cover all $1$-entries of $U$, there must exist an $i \in [d]$
such that $(x,y) \in B_i$.
Therefore, $(x,y)$ is compatible with $S_i$. Furthermore, $(x,y)$ is not influential with respect to the subset $S_i$ (otherwise, we would have added it to $S_i$).
Thus,  entry $(x,y)$ cannot be beneficial for $S$.
\EPF

\subsection{Skeletons of matrices far from 
                    rank $d$ have many beneficial entries}
\label{subsec:matrix-far-many-beneficial}

In this subsection we show that if the matrix $M$ is $\epsilon$-far from 
Boolean rank at most $d$,
then every skeleton has many beneficial entries. To be precise, we prove the contrapositive statement:

\BL\label{lem:few-benefical-matrix-close}
Let $S = \{S_1,\dots,S_d\}$ be a skeleton for $M$ with at most $\frac{\eps^2}{64} n^2$ beneficial entries.
Then $M$ is $\epsilon$-close to Boolean rank $d$.
\EL

In order to prove Lemma~\ref{lem:few-benefical-matrix-close}, we first show how to modify $M$ in at most $\epsilon n^2$ entries,
and then prove that after this modification the resulting matrix $M'$ has Boolean rank at most $d$.
We note that in  all that follows, the reference to beneficial entries is with respect to the given skeleton $S$ stated in Lemma~\ref{lem:few-benefical-matrix-close}.
We start by showing how to modify $M$ using the following {\em Modification rules:}
\begin{enumerate}
\item  \label{mod-rule:all-zero}
Modify each row/column with at least $\epsilon n/8$ beneficial entries to an all-zero row/column.
The number of such rows/columns is at most $\epsilon n/8$. Otherwise, we get more than $\frac{\eps^2}{64} n^2$ beneficial entries.
Therefore, this step accounts for  at most $2n\cdot\epsilon n/8  = \epsilon n^2 /4$  modifications.
\item  \label{mod-rule:few-beneficial}
Modify to $0$'s all beneficial entries  in rows/columns with less than $\epsilon n/8$ beneficial entries.
This accounts for at most $2n\cdot \eps n/8  = \epsilon n^2 /4$ modifications.
\item \label{mod-rule:zero-to-one}
Modify a 0-entry $(x,y)$ to a 1 (where $x$ and $y$ are a row/column with less than $\epsilon n/8$ beneficial entries)
if and only if there exists an $i \in [d]$,
such that row $x$ and column $y$ are both friendly and zero-light with respect to $S_{i}$.

By Claim~\ref{clm:fewzeros}, in this case it holds that
$y \in \Z(x) \setminus \Z(R(S_i))$ and $x \in \Z(y) \setminus \Z(C(S_i))$.
Thus,
the total number of modifications of this type is at most $2n\cdot d \cdot g(\eps,d)n = \eps n^2 /2 $, since $g(\eps,d) = \eps/(4d)$.
\end{enumerate}
Therefore, the total number of modified entries is upper bounded by:
$$\epsilon n^2 /4 +  \epsilon n^2 /4 +  \eps n^2 /2 = \eps n^2.$$

The main issue is hence proving that after this modification, the modified matrix $M'$ has Boolean rank at most $d$.
We first define $d$ subsets $B_1,\dots,B_d$ of $1$-entries, such that each $1$-entry of the modified matrix $M'$ is included in one of these subsets:

  \begin{enumerate}
 \item 
 For each  $(x,y)\in \bigcup_{j=1}^d S_j$ such that $M'[x,y]=1$:
 place $(x,y)$ in $B_i$ for $i \in [d]$ such that $(x,y)\in S_i$.
 \item For each  $(x,y)\notin \bigcup_{j=1}^d S_j$ such that $M'[x,y]=1$:
 place $(x,y)$ in $B_{i}$
 if both row $x$ and column $y$ are friendly and zero-light (in $M$) with respect to $S_i$.

 To verify that such an index $i$ exists for such an entry $(x,y)$, we consider two cases:
     \begin{enumerate}
     \item\label{it:M-x-y-1} $M[x,y]=1$: Since $M'[x,y]=1$ as well, we know that $(x,y)$ is non-beneficial (since beneficial entries were modified to 0).
      By the definition of non-beneficial entries, there exists an index $i$, such that $(x,y)$ is compatible with $S_i$ and non-influential with respect to $S_i$.
      That is, both row $x$ and column $y$ are friendly (in $M$) with $S_i$, and are zero-light (in $M$) with respect to $S_i$.
     \item \label{it:M-x-y-0} $M[x,y]=0$: Since $M'[x,y]=1$, by Modification rule number~\ref{mod-rule:zero-to-one}, such an index $i$ must exist as well.
     \end{enumerate}
 \end{enumerate}

It remains to prove that the subsets $B_1,\dots,B_d$  induce a cover of $M'$ by $d$ monochromatic rectangles.
That is, for each subset $B_i$, every two $1$-entries in $B_i$ are compatible (in $M'$).
We first prove the 
next claim, 
which follows from the modification rules of $M$ and the definition of these subsets.

\BCM
\label{clm:friendly and light}
 If $M'[x,y]=1$ and $(x,y) \in B_{i}$, then row $x$ and column $y$ are friendly and zero-light in $M$  with respect to  $S_{i}$.
\ECM
\BPF
Consider the following cases:
\begin{itemize}
\item
 $(x,y)\in \bigcup_{j=1}^d S_j$:
 Thus, $(x,y) \in S_i$ and by the definition of the skeletons this means that $(x,y)$ is compatible with all entries in $S_i$.
 Hence, row $x$ and column $y$ are friendly with $S_i$.
Furthermore,
given that $(x,y) \in S_i$, we have that $\Z(x) \setminus \Z(R(S_i)) = \emptyset$ and $\Z(y) \setminus \Z(C(S_i)) = \emptyset$,
so that  row $x$ and column $y$ are zero-light with respect to $S_i$.
 \item
 $(x,y)\notin \bigcup_{j=1}^d S_j$ and $M[x,y]=1$:
This case corresponds to Case~\ref{it:M-x-y-1} in the definition of the subsets $B_i$, and so
row $x$ and column $y$ are friendly and zero-light in $M$  with respect to  $S_{i}$ by the definition.
 \item
 $(x,y)\notin \bigcup_{j=1}^d S_j$ and $M[x,y]=0$:
This case corresponds to Case~\ref{it:M-x-y-0} in the definition of the subsets $B_i$, and so
row $x$ and column $y$ are friendly and zero-light in $M$  with respect to  $S_{i}$ by the definition.
\end{itemize}
Since every pair $(x,y)$ fits one of the above cases, the claim follows.
\EPF\\

The next claim concludes the proof that $M'$ has Boolean rank at most $d$, thus 
establishing the proof of Lemma~\ref{lem:few-benefical-matrix-close}.
\BCM\label{clm:Bi-compatible}
For every 
$i \in [d]$, every two entries in $B_i$ are compatible in $M'$.
\ECM
\BPF
Consider any pair of entries $(x_1,y_1),(x_2,y_2) \in B_i$.
By Claim~\ref{clm:friendly and light}, rows $x_1$ and $x_2$ and columns $y_1$ and $y_2$, are friendly and zero-light in $M$
with respect to $S_{i}$.
Furthermore, these rows/columns were not modified by Modification rule number~\ref{mod-rule:all-zero}.

We now show that $M'[x_1,y_2]=1$, where a similar proof holds for $(x_2,y_1)$.
We consider the following cases:
\begin{itemize}
\item
$M[x_1,y_2]= 0$: Since rows $x_1$ and $x_2$ and columns $y_1$ and $y_2$ are friendly and zero-light with respect to $S_{i}$,
then by Modification rule number~\ref{mod-rule:zero-to-one}
 we have $M'[x_1,y_2]= 1$.
\item
$M[x_1,y_2]= 1$: Since rows $x_1$ and $x_2$ and columns $y_1$ and $y_2$ are friendly and zero-light with respect to $S_{i}$, then $(x_1,y_2)$ cannot be
influential with respect to $S_i$.
It remains to show that $(x_1,y_2)$ is compatible with $S_i$, and therefore cannot be beneficial, and thus, was not modified to a $0$
by Modification rule number~\ref{mod-rule:few-beneficial}.

Let $(x',y')\in S_i$. Since row $x_1$  and column $y_2$ are friendly  with respect to $S_{i}$, then $M[x',y_2] = 1$ and $M[x_1,y'] = 1$.
Therefore, $(x_1,y_2)$ is compatible with $S_i$.
\end{itemize}
Claim~\ref{clm:Bi-compatible} follows.
\EPF

\subsection{A sampling lemma}
\label{subsec:sampling}

Before we state and prove the main lemma of this subsection, we first
establish a bound on the size of each of the subsets in a skeleton.

\BCM\label{clm:skel-size}
Let $S =\{S_1,\ldots,S_d\}$ be a skeleton for $M$. Then $|S_i| \leq 8d/\epsilon$
for every 
$i\in [d]$.
\ECM
\BPF
Every entry $(x,y)$ that is added inductively to subset $S_i$ of the skeleton $S$, adds at least $\frac{\eps}{4d}\cdot n$ columns to $Z(R(S_i))$ or rows to $Z(C(S_i))$.
Thus, at most $8d/\epsilon$ entries can be added to $S_i$ until there are no more influential entries with respect to $S_i$.
\EPF\\


\BL\label{lem:submatrix}
Let $0 < \alpha <1$, and suppose that
every skeleton in $\calS(M)$  has at least
$\alpha \cdot n^2$ beneficial entries  in $M$.

Consider selecting, uniformly, independently and at random,
$m = c\cdot \left(\frac{d^2}{\alpha\cdot \eps}\cdot \log \frac{d}{\alpha\cdot \eps}\right)$ entries from $M$
 for a sufficiently large constant $c$, and denoting the subset of selected entries by $U$.
Then with probability at least $2/3$, for every skeleton $S =\{S_1,\ldots,S_d\}\in \calS(M)$  such that $\bigcup_{i=1}^d S_i \subset U$,
there exists a beneficial entry in $U$ for $S$.
\EL
\BPF
Consider selecting $m$ entries from $M$, uniformly, independently and at random, and let $ (x_i,y_i)$
be the $i$'th entry selected, so that each entry $(x_i,y_i)$ is a random variable.
Let $s = 8d/\eps$   and  $m = 200\cdot\frac{d^2}{\alpha\cdot \eps}\cdot \ln \frac{d}{\alpha\cdot \eps}$.

By Claim~\ref{clm:skel-size}, for every skeleton $S =\{S_1,\ldots,S_d\}$ in $\calS(M)$, we have that $|S_i| \leq s$ for every
subset $S_i\in S$. Therefore, $\bigcup_{i=1}^d S_i \leq d \cdot s$.
Observe that for each subset of entries $T$ of size at most $d \cdot s$, the number of skeletons
$\{S_1,\dots,S_d\}$ such that $\bigcup_{i=1}^d S_i = T$ is upper bounded by
 \[
 \left( \sum_{i=0}^s {d\cdot s \choose i}\right)^d \leq \left( (s+1)\cdot {d\cdot s \choose s}\right)^d \;\leq\; (s+1)^d \cdot \left(\frac{e \cdot d\cdot s}{s}\right)^{d\cdot s}
 \; = (s+1)^d \cdot (e\cdot d)^{d\cdot s} \;
 \;.
 \]

For each subset of indices $I \subset [m]$, where $|I| \leq d\cdot s$, suppose that we first select entries $\{(x_i,y_i)\}_{i\in I}$,
and let $T_I$ be the resulting set of entries. By the premise of the lemma,
for each skeleton $S = \{S_1,\ldots,S_d\}\in \calS(M)$ such that $\bigcup_{i=1}^d S_i=T_I$, there are
 at least $\alpha \cdot n^2$ beneficial entries  in $M$.

For our choice of $m$, we have that $m-s > m/2$.
Therefore, if we now select the remaining entries $\{(x_i,y_i)\}_{i\in [m]\setminus I}$,
the probability that we do not obtain any entry that is beneficial for $S$ is at most
$$(1-\alpha)^{m/2} < e^{-\alpha m/2}.$$

By taking a union bound over all subsets $I$ of size at most $d\cdot s$, and all skeletons $S$ such that $\bigcup_{i=1}^d S_i=T_I$,
we get that the probability that there exists a skeleton $S =\{S_1,\ldots,S_d\} \in \calS(M)$
such that $\bigcup_{i=1}^d S_i \subset U$, and   there is no beneficial entry in $U$ for $S$, is upper bounded by
\[
m^{d\cdot s} \cdot (s+1)^d \cdot (e\cdot d)^{d\cdot s} \cdot e^{-\alpha m/2} =
e^{\left(d\cdot s\ln m + d\ln (s+1) + d\cdot s\cdot \ln(e\cdot d)- \alpha m/2\right)} \leq e^{-2} \leq \frac{1}{3}\;
\]
where the first inequality holds for our setting of $s$ and $m$.
\EPF\\

\subsection{Proof of Theorem~\ref{thm:main}}\label{subsec:thm-main-proof}

We can now complete the proof of Theorem~\ref{thm:main}, which builds on Lemmas~\ref{lem:rank-k-no-beneficial},~\ref{lem:few-benefical-matrix-close} and~\ref{lem:submatrix}.
\\

\BPFOF{Theorem~\ref{thm:main}}
If $M$ has Boolean rank at most $d$, then Algorithm~\ref{alg:Boolean test} always accepts since every submatrix of $M$ has Boolean rank at most $d$.

Assume, therefore, that $M$ is $\eps$-far from Boolean rank at most $d$.
By Lemma~\ref{lem:few-benefical-matrix-close}, for every skeleton in $\calS(M)$
there are at least $\frac{\eps^2}{64} n^2$ beneficial entries in $M$.
Therefore, by Lemma~\ref{lem:submatrix} (applied with $\alpha = \frac{\eps^2}{64}$),  for
$m$ as set in Algorithm~\ref{alg:Boolean test},
with probability at least $2/3$, for
 every skeleton $S =\{S_1,\ldots,S_d\} \in \calS(M)$ such that $ \bigcup_{i=1}^d S_i \subset U$,
there exists a beneficial entry $(x,y)\in U$ for $S$.

But by 
Lemma~\ref{lem:rank-k-no-beneficial},
if the Boolean rank of $W$ was at most $d$, then for every $U \subseteq W$, there must exist a skeleton
$S =\{S_1,\ldots,S_d\}  \in \calS(M)$, where $\bigcup_{i=1}^d S_i \subset U$, with no beneficial entries in $U$.
Hence, the Boolean rank of $W$ must be larger than $d$, and thus Algorithm~\ref{alg:Boolean test} will reject as required.
\EPFOF


\section{Testing the binary rank}
\label{sec:binary}

We present simple testing algorithms for the binary rank whose query complexity is exponential in $d$.
Although the query complexity is exponential in $d$, it is strictly smaller than that of the
algorithm derived from the result of~\cite{alon2007efficient} described in Section~\ref{sec:Alon}.
We first give a non-adaptive algorithm whose query complexity is
$O(2^{2d}/\epsilon^2)$, and then use its analysis to design an adaptive algorithm whose query complexity is $O(2^{2d}/\epsilon)$.
We note that variants of these algorithms are also applicable to the Boolean rank.

\subsection{A non-adaptive property testing algorithm for the binary rank}

\bigskip
\fbox{
\begin{minipage}{5.4in}
\BA{{\sf(Test $M$ for binary rank $d$, given $d$ and $\epsilon$ -- non-adaptive version)}}
\label{alg:binary-non-adaptive}
\begin{enumerate}
\item
Select uniformly, independently and at random
$m = 24(2^d+1)/\epsilon$ entries from $M$.
\item
Let $U$ be the subset of entries selected 
and consider the submatrix $W$ of $M$ induced by $U$.
\item
If $W$ has binary rank at most $d$, then accept. Otherwise, reject.
\end{enumerate}

\EA
\end{minipage}
}
\bigskip

The query complexity of the algorithm is  $O(2^{2d}/\epsilon^2)$, and it
always accepts a matrix $M$ that has binary rank at most $d$,
as every submatrix of $M$ has binary rank at most $d$.
Hence, it remains to prove the following lemma:


\BL\label{lem:binary-non-adaptive}
Let $M$ be  a matrix that is  $\epsilon$-far from binary rank at most $d$.
Then Algorithm~\ref{alg:binary-non-adaptive} rejects with probability at least $2/3$.
\EL

In order to prove Lemma~\ref{lem:binary-non-adaptive}, we first establish a couple of claims.
The first is a simple claim regarding the number of distinct rows and columns in matrices with rank at most $d$.

\BCM
\label{clm:size of matrix}
Let $W$ be a $(0,1)$-matrix of binary (or Boolean) rank at most $d$.
Then every submatrix of $W$ has at most $2^d$ distinct rows and at most $2^d$ distinct columns.
\ECM
\BPF
If $W$ has binary rank at most $d$, it clearly has Boolean rank at most $d$. Thus, it suffices to prove the claim for the latter case.
If $W$ has Boolean rank $d$, then the $1$-entries can be covered by $d$ monochromatic rectangles.
Any two rows that share a monochromatic rectangle must have $1$-entries in the columns that belong to this rectangle.
Therefore, there are at most $2^d$ distinct rows in $W$ according to the monochromatic rectangles to which each row can belong.
A similar argument holds for the columns.
\EPF\\

In order to state our next claim, we introduce a few definitions.

\BD[Number of Distinct rows/columns]
\label{distinct}
Denote by $N(R(W))$ the number of {\sf distinct} rows in a submatrix $W$,
and by $N(C(W))$ the number of distinct columns in $W$.
\ED

\BD[New row/column]
A row index $x\in [n]$ is said to be {\sf new} with respect to a submatrix $W$ of $M$,
if by extending $W$ with $x$ we obtain a row different from all current rows of $W$.
That is, if $W$ is the submatrix induced by $(x_1,y_1),\dots,(x_t,y_t)$, then
 $(M[x,y_1],\dots,M[x,y_{t}]) \neq ( M[x_i,y_1],\dots,M[x_i,y_{t}])$ for all $i \in [t]$.
A  new column index $y$ is defined similarly.
\ED

\begin{figure}[htb!]
\captionsetup{width=0.9\textwidth}
\centering
    \includegraphics[width=0.3\textwidth]{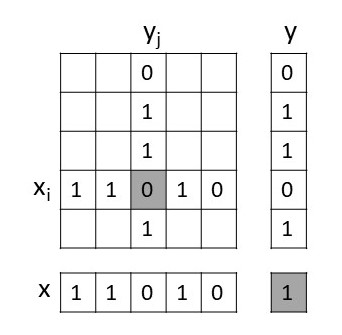}
\caption{\small An illustration for Definition~\ref{def:new corner} (new corner entry).}
\label{fig:corner}
\end{figure}

\BD[New corner entry]
\label{def:new corner}
Let $W$ be a submatrix of $M$ that  is  induced  by entries $(x_1,y_1),\dots,(x_t,y_t)$,
and let $(x,y) \in [n]\times [n]$ be an entry, such that neither $x$ nor $y$ is new for  $W$.
Then $(x,y)$ is said to be a {\sf new corner entry} with respect to $W$ if
there exist $i,j \in [t]$, such that:
\begin{enumerate}
\item
 $(M[x,y_1],\dots,M[x,y_{t}]) = (M[x_i,y_1],\dots,M[x_i,y_{t}])$,
 \item
 $( M[x_1,y],\dots,M[x_{t},y]) = ( M[x_1,y_j],\dots,M[x_{t},y_j])$,
 \item
 $M[x,y] \neq M[x_i,y_j]$.
 \end{enumerate}
\ED
For an illustration of a new corner entry, see Figure~\ref{fig:corner}.

\BCM
\label{claim:new}
Let $W$ be a submatrix of $M$  induced by $(x_1,y_1),\dots,(x_t,y_t)$.
If  the binary rank of $W$ is at most $d$, and $M$ is $\eps$-far from binary rank at most $d$,
then one of the following must hold:
\begin{enumerate}
\item
The number of  row indices $x\in [n]$ that are new with respect to $W$ is greater than $(\eps/3)n$;
\item
The number of  column indices $y\in [n]$ that are new with respect to $W$ is greater than $(\eps/3)n$;
\item
The number of  corner entries $(x,y)\in [n]\times [n]$ that are new with respect $W$  is greater than $(\eps/3)n^2$.
\end{enumerate}
\ECM
\BPF
Assume, contrary to the claim, that none of the three statements stated in the claim holds.
In such a case, we can modify $M$ as follows, and obtain a matrix $M'$, which we shall show
has binary rank at most $d$:
\begin{itemize}
\item
For each row index $x\in [n]$ that is new with respect to $W$, row $x$ in $M'$ is set to be the all-zero row.
\item
For each column index $y\in [n]$ that is new with respect to $W$, column $y$ in $M'$ is set to be the all-zero column.
\item
For each entry $(x,y)\in [n]\times [n]$ that is a new corner entry with respect to $W$: Let $i,j \in [t]$ be such that  $(M[x,y_1],\dots,M[x,y_{t}]) = (M[x_i,y_1],\dots,M[x_i,y_{t}])$
and $( M[x_1,y],\dots,M[x_{t},y]) = ( M[x_1,y_j],\dots,M[x_{t},y_j])$.
Set  $M'[x,y] = M[x_i,y_j]$.
\item All other entries of $M'$ are as in $M$.
\end{itemize}
Observe that by the above modification rules,
 for every entry $(x,y)$ such that neither $x$ nor $y$ is new with respect to $W$,
there exist indices $i,j \in [t]$ as specified in the third item above, and it holds that $M'[x,y] = M[x_i,y_j]$.

By the premise of the claim, the number of entries that $M'$ and   $M$ differ on, is at most
$2\cdot (\eps/3)n\cdot n + (\eps/3)n^2 = \eps n^2$.
As we show next, since $W$ has binary rank at most $d$,
so does the resulting matrix $M'$,
in contradiction to our assumption that $M$ is $\eps$-far from binary rank at most $d$.

To verify that $M'$ has binary rank at most $d$,
consider a partition of the $1$-entries of $W$ into $d' \leq d$ monochromatic rectangles
$B_1,\dots,B_{d'}$.
We shall show how, based on this partition, we can define a partition of all $1$-entries of $M'$ into
$d'$ monochromatic rectangles $B'_1,\dots,B'_{d'}$.

Note that for any $1$-entry $(w,z)$  in $M'$,
neither the row index $w$ is new with respect to $W$ nor the column index $z$ is a new with respect to $W$
(since otherwise, we would have modified row $w$ and/or column $z$ to the all-zero row, and thus, $M'[w,z]=0$).
Therefore, there exists a row index $i(w)$, and column index $j(w)$ such that:
\begin{equation}
\begin{split}\label{wz}
(M[w,y_1],\dots,M[w,y_t]) = (M[x_{i(w)},y_1],\dots,M[x_{i(w)},y_t]), \\
( M[x_1,z],\dots,M[x_{t},z]) = ( M[x_1,y_{j(z)}],\dots,M[x_{t},y_{j(z)}]).
\end{split}
\end{equation}
where $i()$ is a function that maps row $w$ of $M$ to a row $i(w)$ in $W$ as specified in Equation~\eqref{wz},
and if there are several such rows in $W$, then the function $i()$ chooses one arbitrarily. The function  $j()$ is defined similarly for the columns.
Also observe that  if $w = x_s$ for some $s \in [t]$, then $i(w) = s$ and similarly, if $z = y_s$ for some $s\in [t]$, then $j(z) = s$.
Furthermore, as stated above, for such a $1$-entry $(w,z)$ it holds that $M'[w,z] = M[x_{i(w)},y_{j(w)}]$, and therefore, $M[x_{i(w)},y_{j(w)}]=1$.

Now, place $(w,z)$ in $B'_\ell$, where $\ell$ is such that $(x_{i(w)},y_{j(z)}) \in B_\ell$.
In particular, if $(w,z)$ belongs to $W$ and $(w,z) \in B_\ell$, then $(w,z) \in B'_\ell$.

To verify that $B'_1,\dots,B'_{d'}$ is a partition of the $1$-entries of $M'$ into monochromatic rectangles,
consider any pair of $1$-entries in $M'$, $(w,z)$ and $(w',z')$,
such that $(w,z),(w',z') \in  B'_\ell$.
We need to show that $(w,z')$ and $(w',z)$ are also $1$-entries of $M'$, and that
$(w,z'), (w',z) \in B'_\ell$ as well.

\begin{figure}[htb!]
\captionsetup{width=0.9\textwidth}
\centering
    \includegraphics[width=0.4\textwidth]{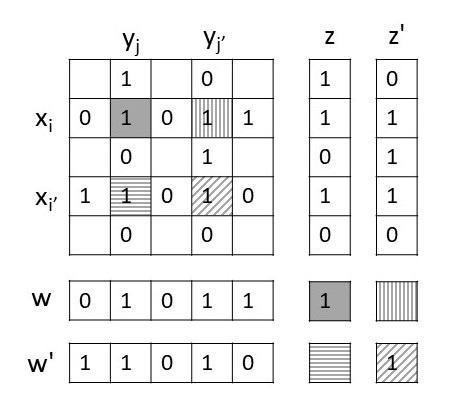}
\caption{\small An illustration of the proof of Claim~\ref{claim:new}.
The figure shows the submatrix $W$, as well as the parts of rows $w,w'$ that are identical to rows $x_{i(w)} = x_i, x_{i(w')} = x_{i'}$ in the submatrix $W$,
and similarly for columns $z,z'$ that are identical to columns $y_{j(z)}= y_j, y_{j(z')} = y_{j'}$ in $W$.
Also shown are entries $(w,z), (w',z'), (w,z'),(w',z)$, where each is filled with the same pattern as the entry it equals to in $W$. If $(w,z)$ and $(w',z')$  belong to $B'_\ell$, then
$(x_i,y_j)$ and $(x_{i'},y_{j'})$ belong  $B_\ell$. This implies that $(x_i,y_{j'})$ and $(x_{i'},y_j)$ also belong
to $B_\ell$, so that $(w,z')$ and $(w',z)$ belong to $B'_\ell$ as well.
 For simplicity not all entries are specified. }
\label{fig:proof}
\end{figure}

Again, since $(w,z')$ and $(w',z)$ do not belong to a row or column that are new with respect to $W$,
we have that $M'[w,z'] = M[x_{i(w)},y_{j(z')}]$ and $M'[w',z] = M[x_{i(w')},y_{j(z)}]$.
But  $(x_{i(w)},y_{j(z)}) \in B_\ell$
and $(x_{i(w')},y_{j(z')}) \in B_\ell$, and thus, we get that $M[x_{i(w)},y_{j(z')}]=1$, $M[x_{i(w')},y_{j(z)}]=1$,
so that $(w,z')$ and $(w',z)$ are also $1$-entries of $M'$.
Furthermore,
$(x_{i(w)},y_{j(z')}) \in B_\ell$ and $(x_{i(w')},y_{j(z)})\in B_\ell$ (by the definition of $B_1,\dots,B_{d'}$), so that
$(w,z')$ and $(w',z)$ both belong to $B'_\ell$,
as required. See Figure~\ref{fig:proof} for an illustration.
\EPF\\

We can now prove Lemma~\ref{lem:binary-non-adaptive}, thus completing the proof of correctness of Algorithm~\ref{alg:binary-non-adaptive}.\\

\BPFOF{Lemma~\ref{lem:binary-non-adaptive}}
For the sake of the analysis, we consider Algorithm~\ref{alg:binary-non-adaptive} as if it proceeds in $m=O(2^d/\epsilon)$ iterations, where it starts with the empty $0\times 0$ submatrix, $W_0$,
and in each iteration it extends the submatrix it has with a row and a  column whose indices are selected uniformly, independently, at random from $[n]$.

For each $t \in [m]$, let $W_{t-1}$ be the submatrix of $M$ of size $(t-1) \times (t-1)$ that is considered in the beginning of iteration $t$,
and let $x_{t}$ and $y_{t}$ denote, respectively, the indices of the row and column selected  in the $t$'th iteration.
Therefore,  $W_t$ is the submatrix induced by $x_1,\dots,x_t$ and $y_1,\dots,y_t$.
We shall show that with probability at least $2/3$, 
the rank of the final submatrix, $W_m$, is greater than $d$, and therefore, the algorithm will reject as required.



%
We know by Claim~\ref{clm:size of matrix}, that if $M$ has binary rank at most $d$, then for any submatrix $W$ of $M$
it holds that  $N(R(W)) \leq 2^d$ and  $N(C(W)) \leq 2^d$.
Hence, if  
either $N(R(W_{t})) > 2^d$ or  $N(C(W_{t})) > 2^d$,
then the rank of $W_t$ is greater than $d$, so that the algorithm will certainly reject.
It is of course possible that $W_t$ has binary rank greater than $d$, although both
$N(R(W_t)) \leq 2^d$ and  $N(C(W_t)) \leq 2^d$, and in this case, the algorithm  rejects as well.

Therefore, for any $t < m$, if the binary rank of $W_{t-1}$ is at most $d$, and given that
in iteration $t$, entry $(x_t,y_t)$ is selected  uniformly, independently at random, then by Claim~\ref{claim:new},
with probability at least $\epsilon/3$, either:
\begin{itemize}
\item
 $x_t$  is new for $W_{t-1}$, so that $N(R(W_t)) > N(R(W_{t-1})$,
 \item
  or $y_t$ is new for $W_{t-1}$, so that $N(C(W_{t})) > N(C(W_{t-1})$,
 \item
or $(x_{t},y_{t})$  is a new corner entry for $W_{t-1}$.
In this case, let  $x_i$ and $y_j$ be as defined in Definition~\ref{def:new corner}.
Thus,
$( M[x_1,y_{t}],\dots,M[x_{t-1},y_{t}] ) = ( M[x_1,y_j],\dots,M[x_{t-1},y_j])$, and in particular, $M[x_i,y_{t}] = M[x_i,y_j]$.
However, $M[x_{t},y_{t}] \neq M[x_i,y_j]$, implying that $N(R(W_{t})) > N(R(W_{t-1}))$ in this case as well.
A similar argument shows that $N(C(W_{t})) > N(C(W_{t-1}))$.
\end{itemize}
To summarize, if the binary rank of $W_{t-1}$ is at most $d$,
then with probability at least $\epsilon/3$, the number of distinct rows or the number of distinct columns, or both, of $W_{t}$ increases compared to that of $W_{t-1}$.
We thus, have to bound the probability that after all $m = \Theta(2^d/\eps)$ iterations, the number of distinct rows and the number of distinct columns of $W_m$, are both at most $2^d$.

To do so, define for each $t\in [m]$,
a Bernoulli random variable $\chi_t$, where $\chi_t =1$ if and only if
$N(R(W_t)) > N(R(W_{t-1}))$ or $N(C(W_{t})) > N(C(W_{t-1}))$, or both.
While the random variables $\chi_1,\dots,\chi_m$ are not independent,
we have that for any $t \in [m]$:
$$\Pr[\chi_{t}=1 | \mbox{the binary rank of $W_{t-1}$ is at most $d$}] \geq \eps/3.$$

Furthermore, if $\sum_{t=1}^m \chi_t \geq 2^d + (2^d+1)$, then necessarily $\max\{N(R(W_m)) ,N(C(W_m))\} > 2^d$,
so that the binary rank of $W_m$ is greater than $d$, and the algorithm rejects.
Note that it is possible that the binary rank of $W_m$ is greater than $d$ although
$\max\{N(R(W_m)) ,N(C(W_m))\} \leq 2^d$, and it is possible
 that $\max\{N(R(W_m)) ,N(C(W_m))\} > 2^d$ although $\sum_{t=1}^m \chi_t < 2^d + (2^d+1)$,
 but in either case the algorithm  rejects.

Finally, we show that  
for $m = 24\cdot 2^d/\eps$, with probability at least $2/3$, the binary rank of  $W_m$, is greater than $d$.
To this end we define $m$ independent random variables,  $\tilde{\chi}_1,\dots,\tilde{\chi}_m$,
where $\Pr[\tilde{\chi}_t =1] = \eps/3$, so that:
$$\Pr\left[\mbox{the binary rank of $W_m$ is at most $d$} \right] \leq \Pr\left[\sum_{t=1}^m \tilde{\chi}_t \leq 2^{d+1}\right].$$
By applying a multiplicative Chernoff bound, given the setting of $m = 24\cdot 2^d/\eps$, so that the expected value of
$\sum_{t=1}^m \tilde{\chi}_t$ is greater than $2 \cdot 2^{d+1}$, we get that:
$$\Pr\left[\sum_{t=1}^m \tilde{\chi}_t \leq 2^{d+1}\right] \leq \exp(-(m/3)(1/2)^2) < 1/3,$$
and the lemma is established.
\EPFOF

\subsection{An adaptive property testing algorithm for the binary rank}

We next describe an adaptive algorithm for the binary rank whose query complexity is $O(2^{2d}/\epsilon)$.
The idea is simple: we modify Algorithm~\ref{alg:binary-non-adaptive} so that it is  ``closer'' to the analysis described in its proof of correctness.
Namely, the modified algorithm works  in $m$ iterations, where $m=\Theta(2^d/\eps)$ is as set in Algorithm~\ref{alg:binary-non-adaptive}.
In each iteration it maintains a submatrix $W_{t-1}$ of $M$, and selects a random row index $x_{t}$  and a random column index $y_{t}$.
It 
extends $W_{t-1}$ 
by $x_{t}$ and $y_{t}$ only if this increases the number of distinct rows/columns in the submatrix.
Therefore, the number of rows/columns of each submatrix $W_t$ never exceeds $2^d$, and thus the total number of queries is bounded by $O(m\cdot 2^d) = O(2^{2d}/\epsilon)$.
Specifically, the modified algorithm is as follows:

\bigskip\noindent
\begin{center}
\fbox{
\begin{minipage}{5.4in}
\BA{{\sf(Test $M$ for binary rank $d$, given $d$ and $\epsilon$ -- adaptive version)}}
\label{alg:binary-adaptive}
\begin{enumerate}
\item
Set $W_0$ to be the empty $0 \times 0$ matrix.
\item
for $t = 1$ to $m$:
\begin{enumerate}
\item
Select, uniformly, independently and at random,  an index $x_{t} \in [n]$ and an index $y_{t} \in [n]$.
\item
Consider the matrix $W'_{t-1}$ obtained by extending $W_{t-1}$ with $x_{t}$ and $y_{t}$.
If
$$\max\{N(R(W'_{t-1})) ,N(C(W'_{t-1}))\} > \max\{N(R(W_{t-1})) ,N(C(W_{t-1}))\}$$
then set $W_{t} = W'_{t-1}$. Otherwise, $W_{t} = W_{t-1}$.
\item
If the binary rank of $W_{t}$ is greater than $d$, then stop and reject.
\item
If no step resulted in a rejection then accept.
\end{enumerate}
\end{enumerate}
\EA
\end{minipage}
}
\end{center}

\vspace{1ex}
Similarly to the non-adaptive algorithm (Algorithm~\ref{alg:binary-non-adaptive}),
if $M$ has binary rank at most $d$, then Algorithm~\ref{alg:binary-adaptive} always accepts.
On the other hand, the argument given in the proof of Lemma~\ref{lem:binary-non-adaptive}
directly implies that if $M$ is $\eps$-far from binary rank at most $d$, then
Algorithm~\ref{alg:binary-adaptive} rejects with probability at least $2/3$.

\section{Testing the  rank using Alon, Fischer and Newman~\cite{alon2007efficient}}
\label{sec:Alon}

For a finite collection $F$ of $(0,1)$-matrices, denote by $\mathcal{P}_F$
the set of all $(0,1)$-matrices that do not contain as a submatrix any row
and/or column permutation of a member of $F$.
The following is proved in~\cite{alon2007efficient}.

\BT[{\cite[Thm. 1.4, Cor. 6.4]{alon2007efficient}}]
\label{Alon complexity}
Let $F$ be a finite collection of $k\times k$ or smaller $(0,1)$-matrices.
There is  a non-adaptive one-sided error testing algorithm for $\mathcal{P}_F$, whose query complexity is
$(\frac{k}{\epsilon})^{O(k^4)}$,  and whose running time  is polynomial in its query complexity.
\ET

In what follows we describe how to use this result to design a testing algorithm for the set
$\mathcal{S}_d$ of all $(0,1)$-matrices of Boolean rank at most $d$.
An analogous result applies to the binary rank.

We must first define a family $F$, such that $\mathcal{P}_F = \mathcal{S}_d$.
Let $F_{d}$ be the set of all $(0,1)$-matrices of Boolean rank  $d$, without
repetitions of rows or columns in the matrices in $F_d$. We next show that
$\mathcal{P}_{F_{d+1}} = \mathcal{S}_{d}$

\BCM
\label{clm:submatrix}
Let $A$ be a matrix of Boolean rank $d$. Then it has a submatrix of Boolean
rank $i$ for every $ 0 < i < d$.
\ECM
\BPF
This follows from the fact that if $B$ is some submatrix of $A$, then
extending $B$ by a row or column increases its  Boolean rank by at most $1$.
\EPF

\BCM\label{clm:PFd-S-d-1}
$\mathcal{P}_{F_{d+1}} = \mathcal{S}_{d}$.
\ECM
\BPF
If $A \in \mathcal{S}_{d}$, then it cannot contain a submatrix of Boolean rank $d+1$,
and, therefore, $A\in \mathcal{P}_{F_{d+1}}$.

Assume now that $A \not \in \mathcal{S}_{d}$. Therefore, $A$ must contain a submatrix $B$ of Boolean rank at least $d+1$.
But then by Claim~\ref{clm:submatrix}, $B$ contains a submatrix $C$ of Boolean rank exactly $d+1$ that is also a submatrix of $A$,
and, therefore, $A \not \in \mathcal{P}_{F_{d+1}}$.
\EPF\\

Using Claim~\ref{clm:PFd-S-d-1},
as well as Claim~\ref{clm:size of matrix} and Theorem~\ref{Alon complexity}, we get:
\begin{coro}
\label{coro:alon}
There exists a property testing algorithm for $\mathcal{P}_{F_{d+1}}= \mathcal{S}_{d}$ whose query complexity is
$(\frac{2^d}{\epsilon})^{O(2^{4d})}$.
\end{coro}

The question is what is the smallest $k = k(d)$ such that
all the matrices in $F_{d}$ are of size at most $k \times k$.
Claim~\ref{clm:size of matrix} implies that $k \leq 2^{d}$,
and it is clear that $k \geq d$, since matrices of smaller size have rank less than $d$.
If  $k = o(2^d)$ then
the upper bound in Corollary~\ref{coro:alon} can be improved,
or it may be the case that $k$ is indeed exponential in $d$, and the complexity we get using this approach is tight.

\subsection*{Acknowledgments}
The second author would like to acknowledge the support of the
 Israel Science Foundation (grant No.~1146/18) and the Kadar family award.

\bibliographystyle{plain}
\bibliography{biclique}

\end{document}

%% file: pream.tex
\newtheorem{theo}{Theorem}

\newtheorem{lem}{Lemma}
\newtheorem{claim}[lem]{Claim}

\newtheorem{coro}[lem]{Corollary}

\newtheorem{define}{Definition}

\newtheorem{alg}{Algorithm}
\newtheorem{proc}{Procedure}

\newcommand{\BE}{\begin{enumerate}} \newcommand{\EE}{\end{enumerate}}
\newcommand{\BI}{\begin{itemize}} \newcommand{\EI}{\end{itemize}}
\newcommand{\BDes}{\begin{description}}\newcommand{\EDes}{\end{description}
}
\newcommand{\BT}{\begin{theo}} \newcommand{\ET}{\end{theo}}
\newcommand{\BL}{\begin{lem}} \newcommand{\EL}{\end{lem}}
\newcommand{\BD}{\begin{define}} \newcommand{\ED}{\end{define}}
\newcommand{\BCM}{\begin{claim}} \newcommand{\ECM}{\end{claim}}
\newcommand{\BC}{\begin{coro}} \newcommand{\EC}{\end{coro}}
\newcommand{\BA}{\begin{alg}} \newcommand{\EA}{\end{alg}}
\newcommand{\BP}{\begin{proc}} \newcommand{\EP}{\end{proc}}

\def\FullBox{\hbox{\vrule width 8pt height 8pt depth 0pt}}
\newcommand{\qed}{\;\;\;\FullBox}
\newenvironment{proof}{\noindent{\bf Proof:~~}}{\(\qed\)}
\newcommand{\BPF}{\begin{proof}} \newcommand {\EPF}{\end{proof}}
\newenvironment{proofof}[1]{\noindent{\bf Proof of {#1}:~~}}{\(\qed\)}
\newcommand{\BPFOF}{\begin{proofof}} \newcommand {\EPFOF}{\end{proofof}}

\newcommand{\BEQ}{\begin{equation}} \newcommand{\EEQ}{\end{equation}}
\newcommand{\BEQN}{\begin{eqnarray}}\newcommand{\EEQN}{\end{eqnarray}}

\newlength{\saveparindent}
\newlength{\saveparskip}
\renewcommand{\Pr}{{\rm Pr}}

\newcommand{\eqdef}{\stackrel{\rm def}{=}}

\newcommand{\eps}{\epsilon}

\newcommand{\calS}{{\cal S}}